# RADIO EMISSION OF THE MOON
# BEFORE AND AFTER
# THE LUNAR PROSPECTOR IMPACT

Berezhnoi A.A.*, Gusev S.G.**, Khavroshkin O.B.***,
Poperechenko B.A.**, Shevchenko V.V.*, Tzyplakov V.A.***

* Sternberg Astronomical Institute,
Universitetskij pr., 13, 117234 Moscow, Russia
Tel. 007-095-939-1029/Fax 007-095-932-88-41
e-mail ber@sai.msu.ru/ shev@sai.msu.ru

** Special Design Office, Moscow Energetic Institute,
Krasnokazarmennaia 14, 111250 Moscow Russia
Fax: +7 095 3625576

*** 123810 Schmidt Institute of Physics
of the Earth of Russian Academy of Sciences,
B.Gruzinskaya 10, Moscow D242, Russia
Fax: +7 095 2549088. e-mail: khavole@uipe-ras.scgis.ru

ABSTRACT

The direct impact of the American spacecraft Lunar Prospector into the Moon in the south polar regions occurred in July 31, 1999. This collision was accompanied by seismic effects comparable to a big moonquake. Due to the peculiarities of the lunar lithosphere structure the duration of seismic effects on the Moon is some hours and such impact events may be accompanied by radioseismic radiation (RSR) of the lunar surface layers. Radioseismic radiation during an earthquake has been detected (Sobolev, Demin, 1980). The registration of the lunar RSR resulting from moonquakes or meteoroid impacts may lead to the creation of a new information channel for investigation of lunar seismic activity and the internal structure.

The kinetic energy of the Lunar Prospector impact into the Moon was about $10^{15}$ ergs. In this case the energy of the lunar RSR may be equal to $10^8$ - $10^{10}$ ergs. Unfortunately the uncertainty in estimates of the radiation duration and the frequency dependence of emission power is high.

For the registration of lunar RSR after the Lunar Prospector impact radio observations of the Moon on the two 64-m radio telescopes of the SDO MEI at Medvezh'i Ozera and Kaliazin were conducted during July 30 - August 2, 1999. Due to strong radio noise the quality of radio observations at Medvezh'i Ozera was not good. The radio observations of the lunar polar regions, seismically active and seismically passive regions were conducted on the Kaliazin radio telescope at 13 and 21 cm. The sensitivity of the radio receiver was 90 Jy at both radio wavelengths, and the accuracy of the measurements of the lunar radio emission was 1-3 K. Changes of the radioemission of the south polar regions by comparison with other regions were detected. The Lunar Prospector impact may have been the cause of these changes.

Recommendations about the registration of the Lunar RSR during Leonid's activity in November 2000 are given.

## INTRODUCTION

For direct detection of water ice on the Moon the American spacecraft Lunar Prospector was collided with the Moon in the south lunar region. This impact was accompanied by seismic effects comparable with a big moonquake. Due to singularities of the lunar lithosphere the duration of such a seismic effect may be several hours and can be accompanied by radioemission of the subsurface lunar regolith. Such effect has been detected on the Earth during earthquakes. The electromagnetic emission of seismic sources on the Earth arises from the motion of electric charges due to the piezoelectric effect during micro deformation and microdestruction of the medium [1]. The nature of such electromagnetic emission has not been investigated very well. Such radioseismic processes probably occur on the Moon and planets of the terrestrial group. Sporadic radioemission can occur during the collision of big meteorites with the Moon; regular radioemission can occur due to tidal forces, temperature gradients and local mechanic stresses. On the Earth we can see a good correlation between the intensity of radioemission and the energetic state of a geologic medium and the microseismic noise field [2-4]. So a change of energetic state of the geologic medium (for example due to a meteoroid impact) can be accompanied by a change of intensity of radioemission. We tried to detect such an effect during the collision between the American spacecraft Lunar Prospector (M~ 200 kg, V~ 1.7 km/s) and the Moon. At what wavelengths and with what bandwidth could we detect an emission? The usual range of wavelengths of radiowaves registered by radio telescopes exceeds the usual wavelengths of radioemission on the Earth by several orders of magnitude [5]. For effective investigation of earth seismic emission we must use narrow band filtration and envelope separation [6 - 8]. So such an approach should also be used for the detection of lunar radioemission of seismic origin. The 64-m radio telescopes at Medvezh'i Ozera and Kaliazin satisfied these requirements.

## PARAMETERS
## OF LUNAR RADIOEMISSON

In accordance with experimental data the power of source of radioemission of seismic origin on the Earth is equal to $10^{-2}$ - $10^{6}$ Wt and depends on the volume of the deforming medium and the density of cracks [1]. On the Earth the maximum of intensity of such emission is observed at 100 kHz - 100 MHz. The modulation period can be from 0.1 s to 20 minutes.

The kinetic energy of the collision between the Lunar Prospector and the Moon was equal to $2*10^{8}$ J. The seismic energy of such an impact can be 1-5 % of the kinetic energy (~ $10^{6}$ J). Due to the low efficiency of transformation of seismic energy to electromagnetic emission the power of radioemission of seismic origin we estimate as 10 - $10^{3}$ J. But the spectrum of such emission can be very wide ($10^{5}$-$10^{10}$ Hz). Due to the high soundness of the lunar regolith the duration of radioemission could be some hours (for example the relaxation time after a big meteoroid impact is 1-5 hours in accordance with Apollo results)

## OBSERVATIONS

We registered the lunar radioemission using the 64-m radio telescopes at Kaliazin at 13 and 21 cm and at Medvezh'i Ozera at 90 cm. Due to strong radio noise the quality of radio observations at Medvezh'i Ozera was not good. The sensitivity of the radio receiver at Kaliazin was 90 Jy in both radio wavelengths. The receivers were equipped with low-noise amplifiers with microcryogenic systems. The bandwidths at 13 and 21 cm are 8' and 12' accordingly, sometimes exceeding the angular size of the Moon (30'). The calibration of radio fluxes was conducted using the standard method used during interferometric observations. The sensitivity of the radiotelescope was 0.3 K, the accuracy of the measurements of the lunar radio emission was 1-3 K. The registration of the signal was conducted by paper recording over four days (July30 - August 2).

We observed three regions on the Moon: the South Polar Region, seismically active and seismically passive regions. You can see results of our observations in the pictures 1, 2. Changes of the radioemission flux after the Lunar Prospector impact were detected in comparison with radioemission before the impact. It is interesting that this effect was detected for all three lunar regions. Also the correlation (q = 0.5) between the variations of effective temperature at different wavelengths are observed that can be evidence of similarity of nature of radioemission at these wavelengths. The maximum temperature variations are characteristic for the South Polar Region. The Lunar Prospector impact may have been the cause of these changes. It is important that such changes are observed only within 9-12 hours of the Lunar Prospector impact.

# CONCLUSIONS

We detected significant changes of lunar radioemission after the Lunar Prospector impact in comparison with radioemission before this impact.

A special calibration system is needed for the next observations of lunar radioemission of seismic origin.

We can detect Lunar RSR during the Leonid activity on November 17, 2000.

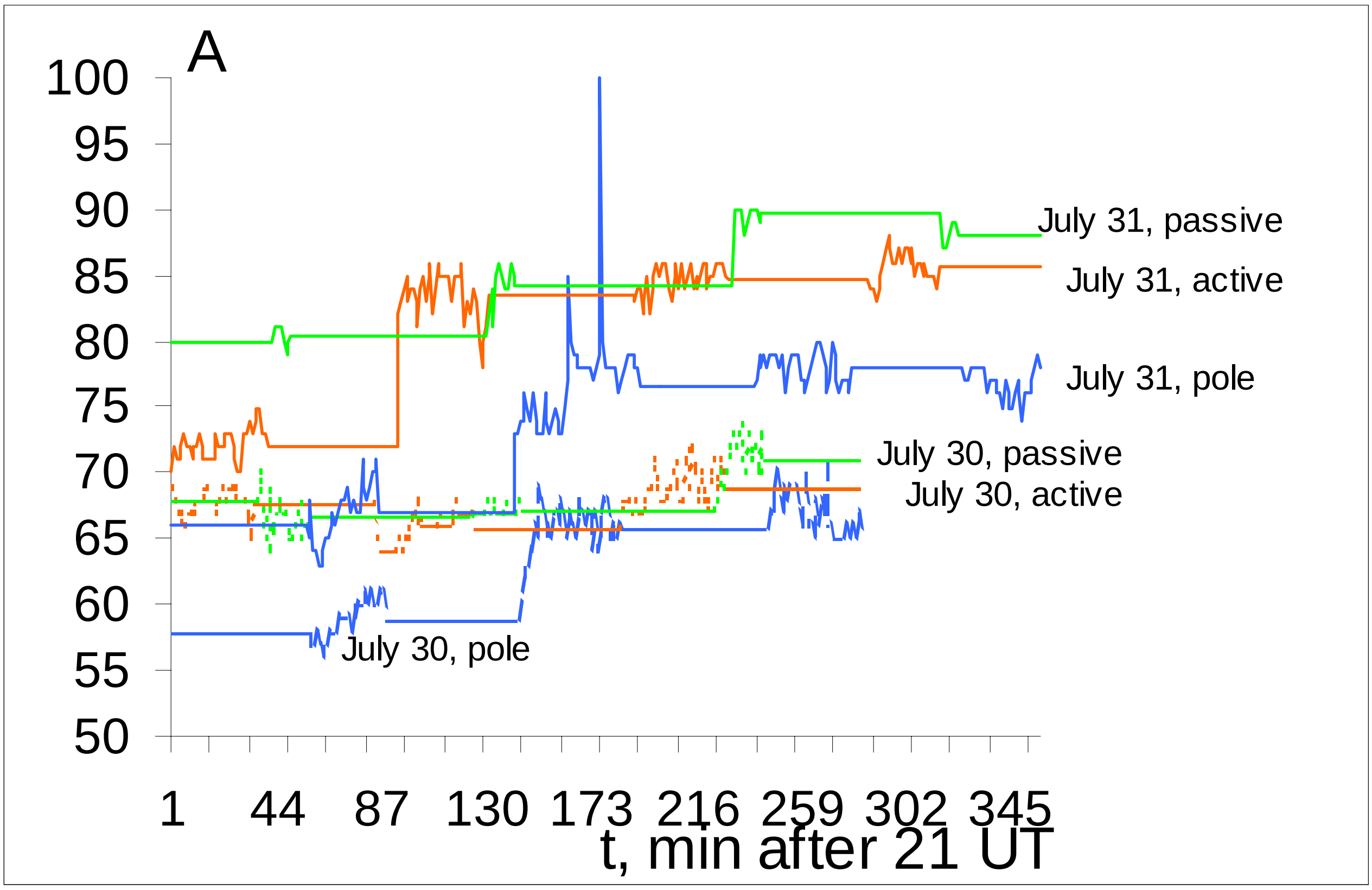

Figure 1. Results of observations of lunar radioemission from the seismically active, seismically passive, and south polar regions at 13 cm before and after the Lunar Prospector impact. Effective temperature T ~ (5-6)* A, where A is the amplitude of signal.

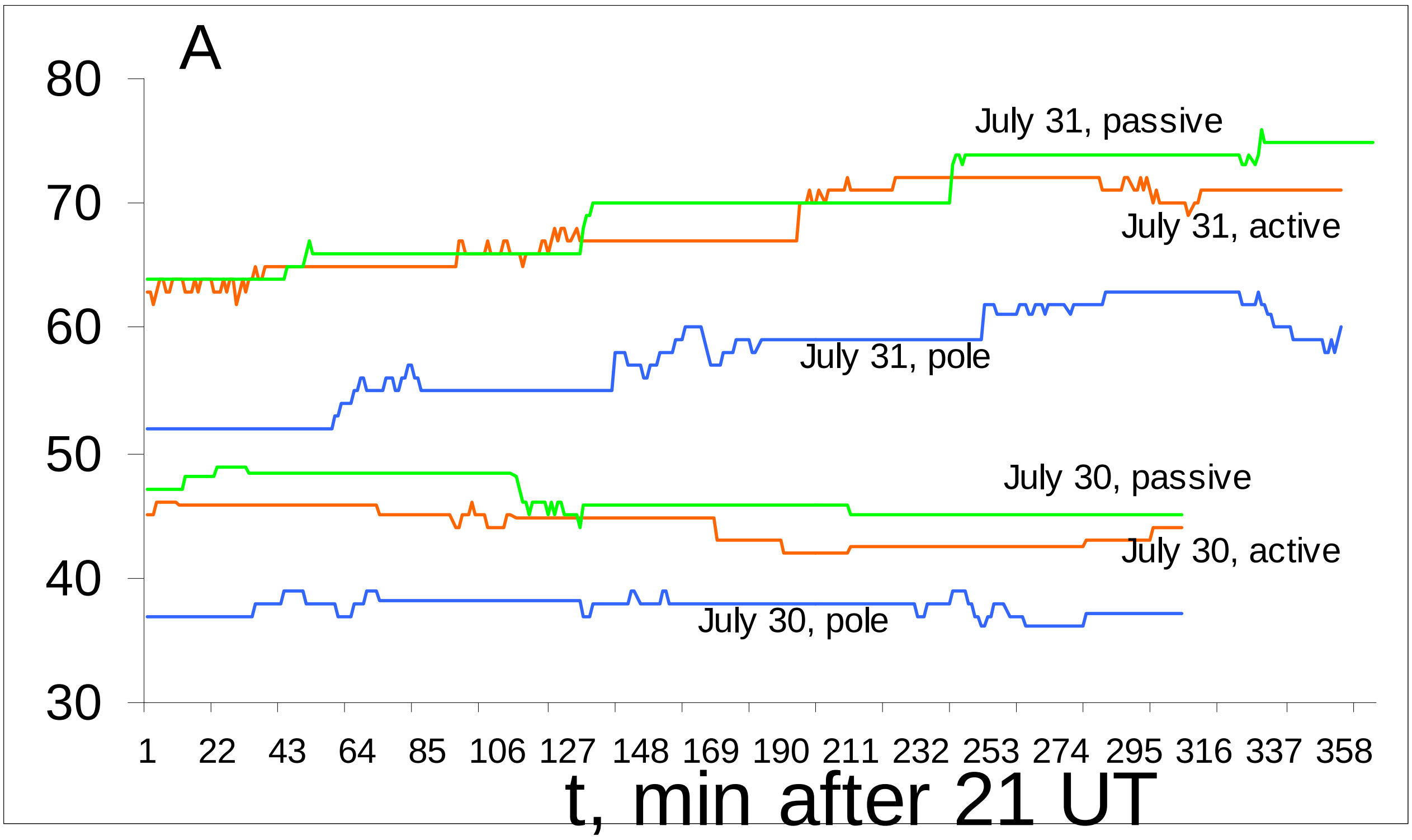

Figure 2. Results of observations of lunar radioemission from the seismically active, seismically passive, and south polar regions at 21 cm before and after the Lunar Prospector impact. Effective temperature T ~ (5-6)* A, where A is the amplitude of signal.